\documentstyle[12pt,epsfig]{article}
\begin{document}
\pagestyle{empty}
\renewcommand{\thefootnote}{\fnsymbol{footnote}}
\def\lsim{\raise0.3ex\hbox{$<$\kern-0.75em\raise-1.1ex\hbox{$\sim$}}}
\def\gsim{\raise0.3ex\hbox{$>$\kern-0.75em\raise-1.1ex\hbox{$\sim$}}}
\def\noi{\noindent}
\def\nn{\nonumber}
\def\bea{\begin{eqnarray}}  \def\eea{\end{eqnarray}}
\def\beq{\begin{equation}}   \def\eeq{\end{equation}}
\def\beeq{\begin{eqnarray}} \def\eeeq{\end{eqnarray}}
\def\R{ {\rm R \kern -.31cm I \kern .15cm}}
\def\C{ {\rm C \kern -.15cm \vrule width.5pt \kern .12cm}}
\def\Z{ {\rm Z \kern -.27cm \angle \kern .02cm}}
\def\N{ {\rm N \kern -.26cm \vrule width.4pt \kern .10cm}}
\def\1{{\rm 1\mskip-4.5mu l} }
\def\lsim{\raise0.3ex\hbox{$<$\kern-0.75em\raise-1.1ex\hbox{$\sim$}}}
\def\gsim{\raise0.3ex\hbox{$>$\kern-0.75em\raise-1.1ex\hbox{$\sim$}}}
\def\sq{\hbox {\rlap{$\sqcap$}$\sqcup$}}
\oddsidemargin  5pt
\vbox to 2 truecm {}
\centerline{\Large \bf A unitary model for structure functions}
\vskip 3 truemm  
\centerline{\Large \bf and diffractive production at small x}

\vskip 1 truecm
\centerline{\bf A. Capella, E. G. Ferreiro, C. A. Salgado}
\centerline{Laboratoire de Physique Th\'eorique\footnote{Unit\'e Mixte de Recherche -
CNRS - UMR n$^{\circ}$ 8627}}  \centerline{Universit\'e de Paris XI, B\^atiment 210,
F-91405 Orsay Cedex, France}

\vskip 5 truemm
\centerline{\bf A. B. Kaidalov}
\centerline{ITEP, B. Cheremushkinskaya ulitsa 25}
\centerline{117259 Moscou, Russia}
\vskip 1 truecm 
\begin{abstract}
We propose a unified approach which
describes both structure functions in the small-$x$
region and diffractive production in $\gamma^*p$-interactions. It is shown that the model,
based on reggeon calculus and a quark-parton picture of the interaction, gives a good
description of available experimental data in a broad region of $Q^2$  (including $Q^2 =
0$) with a single Pomeron of intercept 
$\alpha_P(0) = 1.2$. Predictions for very small
$x$ are given and the problem of saturation of parton densities is discussed. 
   \end{abstract}

\vskip 1 truecm

\noindent LPT Orsay 00-42 \par
\noindent April 2000 \par

\newpage
\pagestyle{plain}
\baselineskip=24 pt

\section{Introduction}
\hspace*{\parindent} Investigation of high-energy interactions of virtual photons with
nucleons and nuclei gives information on the dynamics of high-density partonic systems as
well as on the transition between perturbative and nonperturbative regimes in QCD.  \par

Experiments at HERA have found two extremely important properties of small-$x$ physics~:
a fast increase of parton densities as $x$ decreases and the existence of substantial
diffractive production in deep inelastic scattering (DIS). \par

From a theoretical point of view there are good reasons to believe that the fast increase
of the $\sigma_{\gamma^*p}$ with energy in the HERA energy range will change to a milder
(logarithmic) increase at much higher energies. This is due to unitarity effects, which
are related to shadowing in highly dense systems of partons - with eventual
``saturation'' of densities. This problem has a long history (for reviews see \cite{1r})
and has been extensively discussed in recent years \cite{2r}. It is closely connected to
the problem of the dynamics of very high-energy heavy ion collisions \cite{3r}. \par

In this paper we will address the problem of unitarization in $\gamma^*p$-interactions
using the reggeon calculus \cite{4r} with a supercritical Pomeron ($\Delta_P \equiv
\alpha_P(0) - 1 >0$). In this approach the unitarization effects mentioned above are
described by multi-Pomeron exchanges. The problem of the Pomeron in QCD is not solved yet.
The calculation based on QCD perturbation theory leads to the so-called BFKL-Pomeron
\cite{5r} and the problem of unitarization for such Pomeron has been considered by many
authors \cite{1r}-\cite{3r} in the leading logarithmic approximation. However, recent
calculations \cite{6r} indicate that NLO corrections are large and modify substantially
the picture of high-energy interactions. In particular the intercept of the leading
Regge pole depends on nonperturbative effects \cite{6r,7r}. \par

Here, we will use a more phenomenological approach and will assume that the Pomeron is a
simple pole with an intercept $\alpha_P(0)=1.2$ determined from the analysis of
high-energy hadronic interaction - when all multi-Pomeron contributions are taken into
account \cite{8r}. \par

In ref. \cite{9r} it was suggested that the increase of the effective intercept of the
Pomeron, $\alpha_{eff} = 1 + \Delta_{eff}$, as $Q^2$ increases from zero to several
GeV$^2$ is mostly due to a decrease of shadowing effects with increasing $Q^2$. A
parametrization of the $Q^2$ dependence of $\Delta_{eff}$ such that
$\Delta_{eff} \approx 0.1$ for $Q^2 \approx 0$ (as in soft hadronic interactions)
and $\Delta_{eff} \approx 0.2$ (our input or bare Pomeron intercept) for
$Q^2$ of the order of a few GeV$^2$, gives a good description of all existing data on
$\gamma^*p$ total cross-sections in the region of $Q^2 \ \lsim \ 5 \div 10$~GeV$^2$
\cite{9r,10r}. At larger $Q^2$, effects due to QCD evolution become important. Using the
above parametrization as initial condition in the QCD evolution equation, allows to
describe the data in the whole region of $Q^2$ studied at HERA \cite{9r,11r}.\par

 In the
reggeon calculus \cite{4r} the amount of rescatterings is closely related to diffractive
production. AGK-cutting rules \cite{12r} allow to calculate the cross-section of inelastic
diffraction if contributions of multi-Pomeron exchanges to the elastic scattering amplitude
are known. 
Thus, it is very important for self-consistency of theoretical models to
describe not only total cross-sections, but, simultaneously, 
inelastic diffraction.
Indeed, in the reggeon calculus the variation of $\Delta_{eff}$ with $Q^2$ is directly
related to the corresponding variation of the ratio of diffractive to total
cross-sections. \par

In this paper we present an explicit model 
which
leads to the pattern of energy behaviour of $\sigma_{\gamma^*p}^{(tot)} (W^2,Q^2)$ for
different $Q^2$ des\-cri\-bed above. Moreover, it allows to describe simultaneously
diffraction production by real and virtual photons. The plan of the paper is as follows.
In Section~2 the model is described. Section~3 contains the expressions of the total
cross-section $\sigma_{\gamma^*p}^{tot}$. Formulae for diffractive production are given in
Section~4. Section~5 is devoted to the description of experimental data on
$\sigma_{\gamma^*p}^{(tot)}(W^2,Q^2)$ and diffraction dissociation of the photon. In
Section~6 we compare our results with results of other authors on this subject, discuss
the problem of ``saturation'' and give predictions for future experiments. The
parameters of the model and their numerical values are given in Appendix~1.  

\section{Description of the model}
\hspace*{\parindent} The amplitude of elastic $\gamma^*p$ scattering at very high energy
can be described in terms of the single Pomeron exchange (Fig.~1a) and multi-Pomeron
exchanges (Figs.~1 b, 1c). The Pomeron exchange contribution to $\sigma_{\gamma^*p}^{tot}$
(or to the structure function $F_2 = {Q^2 \over 4 \pi^2\alpha_{e.m}}
\sigma_{\gamma^*p}^{(tot)}$), increases as $\left ( {1 \over x}\right )^{\Delta}$ at
small $x = {Q^2 \over W^2 + Q^2}$. The contribution of $n$-Pomeron exchange behaves as
$\left ( {1 \over x}\right )^{n\Delta}$. These contributions are important at very
small-$x$ to restore unitarity. A resummation of multi-Pomeron contribution is provided by
Gribov reggeon calculus \cite{4r}. In this approach a two-Pomeron exchange
diagram (Fig.~1b), for example, can be represented as a sum over all intermediate
states (Fig.~2) which can be diffractively produced by a single Pomeron exchange. Thus
the magnitude of the two Pomeron exchange in the elastic amplitude is related to the
cross-section of diffractive production. The multi-Pomeron exchanges (Fig.~1c) are
related by AGK-cutting rules \cite{12r} to shadowing corrections to diffractive
production. \par

Let us first discuss the main mechanisms of diffraction dissociation of a
proton and of a virtual photon. For a proton the main contribution to the diagram of
Fig.~2 is due to the proton intermediate state. The dissociation of the proton will be
taken into account in the quasi-eikonal approximation \cite{13r} 
- which works well in
hadronic interactions. As for a virtual photon, it can diffractively produce vector mesons
(Fig.~3a) (the so-called quasi-elastic processes) and large mass states. For $M^2 \gg
Q^2$, the latter correspond to the triple-Regge diagrams of Fig.~3b with $P$ and
$f$-exchanges in the $t$-channel. Models of this type allow to des\-cri\-be diffractive
dissociation both at $Q^2 = 0$ \cite{g} and at large $Q^2$ \cite{15r}. A simple
quasi-eikonal ansatz for multi-Pomeron exchanges based on this type of model was used in
ref. \cite{16r} for the description of multiparticle production in DIS. However, the
quasi-eikonal approximation for diffractive dis\-so\-cia\-tion of a highly virtual photon is
difficult to interpret in terms of reggeon diagrams - while an explicit calculation of these
diagrams for multi-Pomeron exchanges leads to too many parameters for the matrix of
transition between different channels. 
For these reasons, we adopt here a different
approach, based on the quark-parton picture of high-energy interaction of a real or
virtual photon. \par 

Taking into account that a photon dissociates into a
$q\bar{q}$-pair and this pair can have multiple interactions with the target, we
separate such pairs into two groups~: ``aligned'' or asymmetric configurations,
which will be denoted by $A$ and the rest or symmetric configurations 
denoted by $S$. The $A$-configuration was introduced by Bjorken and Kogut \cite{17r} in
the framework of the parton model in DIS and is characterized by a strongly asymmetric
distribution of the relative longitudinal momenta of $q$ and $\bar{q}$ ($z_q \ll 1$,
$z_{\bar{q}} \to 1$ or viceversa). This separation into $A$ and $S$ components is very
important at large $Q^2$, where the $A$-configuration has a large transverse size and
a large interaction cross-section with the proton, while the $S$-component has small
transverse size ($1/Q$) and a small cross-section. On the other hand the
probability of the $A$-configuration decreases as $1/Q^2$ and both configurations
lead to the same behaviour, $\sigma_{\gamma^*p}^{(tot)} \sim 1/Q^2$, at large $Q^2$
\cite {17r}. What is important, however, is that these two components give rise to
a different behavior for diffraction~: the $A$-component gives the same $1/Q^2$
behaviour for total and diffractive cross-sections, while the $S$-component leads
to a diffractive cross-section which decreases as $(1/Q^2)^2$ at large $Q^2$. These
results are well known and have been extensively discussed in the literature
\cite{18r,19r,20r} in connection with HERA experiments. \par

The $q\bar{q}$-state is described, in general, by non-perturbative dynamics and only for
large $Q^2$ and small size component it is possible to use perturbative QCD. \par

Finally, we stress that at present energies, due to phase space limitations, only diagrams
with none or one triple reggeon interactions are important. In particular large mass
diffraction corresponds to a triple Pomeron interaction diagram. However, to be more
general we resum all fan-type diagrams - of the type shown in Fig.~4. 
This will be discussed in more detail in the next Section.

\section{The total $\gamma^*p$ cross-section}
\hspace*{\parindent} We formulate the expressions of cross-sections in the impact
parameter space and take into account that $A$- and $S$-components are diagonal~:

\beq
\label{1e}
\sigma_{\gamma^*p}^{(tot)}(s, Q^2) = 4 \int d^2b \ \sigma_{\gamma^*p}^{(tot)}(b, s, q^2)
\quad , \eeq 

\beq
\label{2e}
\sigma_{\gamma^*p}^{(tot)}(b,s,Q^2) = \sum_i g_i(Q^2) \sigma_i^{(tot)}(b, s, Q^2) \quad i
= A, S \quad .
 \eeq

\noi Here $g_i(Q^2)$ is the probability of the $i$-th component, which according to the
discussion above, has been chosen in the form

\beq
\label{3e}
g_A(Q^2) = {g_A(0) \over 1 + {Q^2 \over m_A^2}} \quad ; \quad g_S(Q^2) = g_S(0)
\eeq

\noi where $g_A(0)$, and $m_A^2$ are considered as phenomenological free parameters. \par

The cross-sections $\sigma_i^{(tot)}(b, s, q^2)$ are written in the form

\beq
\label{4e}
\sigma_i^{(tot)}(b,s, Q^2) = {1 - \exp (- C \chi_i(b,s, Q^2) \over 2C} \quad ,
\eeq

\beq
\label{5e}
\chi_i(s,b,Q^2) = {\chi_{i0}^P(s,b,Q^2) \over 1 + a \chi_3(s,b,Q^2)} + \chi_{i0}^f (s, b,
Q^2) \quad .
 \eeq


The expressions for the eikonals $\chi_{i0}$ are~:

\beq
\label{6e}
\chi_{i0}^k(s,b,Q^2) = C_i^k {f_i(Q^2) \over \lambda_{0k}^i (Q^2, \xi)} \exp \left (
\Delta_k \xi - {b^2 \over 4 \lambda_{0k}^i} \right ) \eeq

\noi with $k = P, f$ and $i = A, S$ where

\bea
\label{7e}
&&\Delta_k = \alpha_k(0) - 1 \quad ; \quad \xi = \ell n {s + Q^2 \over s_0 + Q^2} \nn \\
&&\lambda_{0k}^i = R_{0ki}^2(Q^2) + \alpha '_k \ \xi \quad .
\eea

\noi The quantity $\xi$ is chosen in such a way as to behave as $\ell n {1 \over x}$ for
large $Q^2$ and as $\ell n {s \over s_0}$ for $Q^2 = 0$. The functions
$f_i(Q^2)$ have been chosen in the form\footnote{In eqs. (\protect{\ref{3e}}) and
(\protect{\ref{8e}}) we used the simplest parametrization of $g_i(Q^2)$ and $f_i(Q^2)$
consistent with our requirements. In general $g_S(Q^2)$ and $f_A(Q^2)$ can depend on
$Q^2$.}

\beq
\label{8e}
f_i(Q^2) = \left \{ \begin{array}{cl} \displaystyle{{1 \over 1 + Q^2/m_S^2}} &, \quad i = S
\\ & \\
1 &, \quad i = A \quad . \end{array}
\right .  \eeq

\noi The parametrization of the radius $R_{0ki}^2$ is given in Appendix 1. The slopes
of the trajectories $\alpha'_P = 0.25$~GeV$^{-2}$ and 
$\alpha '_f = 0.9$~GeV$^{-2}$ were
fixed at their values known from soft hadronic interactions. \par

We turn next to the denominator in the first term of eq. (\ref{5e}). 
Here $a =
g_{pp}^P(0) r_{PPP}/16 \pi$ where $g_{pp}^P(0)$ and $r_{PPP}(0)$ are,
respectively, the Pomeron-proton coupling and the triple Pomeron vertex, both
at $t = 0$. The expression of 
$\chi_3(s,b,Q^2)$ is given in Section 4. 
Note that eqs. (4)-(6) correspond to a quasi-eikonal model \cite{13r}.
For $a=0$, 
its Born term, eq. (5), is a sum of Pomeron and secondary $(f)
$ reggeon
exchanges. The latter is included in order to be able to use the 
model at energies as low as $\sqrt{s} \sim
10$~GeV. The coefficient $C$ takes into account the dissociation of the proton 
and is
taken to be equal to 1.5 \cite{13r}.
To first order in $a$, the Born term, eq. (\ref{5e}), also contains 
the contribution of $PPP$ and $PfP$ triple regge terms (see Fig. 5). 
This can be easily seen by developing the first term of the r. h. s. of eq. (5)
to first order in $a$, and using the expression of $\chi_3$ in eqs. (26) and
(27).
Taken to all orders in
$a$, eq. (\ref{5e}) corresponds to the resummation of all fan-type diagrams of the
type shown in Fig.~4, using the Schwimmer formula \cite{21r}. 
Here, however, each new branching contains not only a Pomeron as in Fig. 4 but
also a secondary $f$-reggeon exchange. These branchings are controlled by the 
parameters $a \gamma_P$ and $a \gamma_f$, 
respectively 
(see eq. (26))\footnote{Eq. (5) is valid when these couplings are the 
same irrespectively of whether 
the branching takes place off a Pomeron or an $f$.}.
As discussed in the last paragraph of Section 2, 
diagrams with more than one branching (i. e. more than one triple reggeon 
interaction) 
are not important at present energies due to phase space limitations.  
\par

Note that the triple Pomeron interaction introduces a large (gluonic) size component in
our $S$-component. Actually, the separation between the two components is washed out at
large $Q^2$ due to QCD evolution. Note also that the second term of eq. (\ref{5e})
($f$-exchange) does not contain a denominator - as does the first one. Such a denominator
would add to the $f$-exchange $fPf$ and $fff$ terms. However, such double reggeon
$fP$ and $ff$ exchanges are already present in eq. (\ref{4e}) (terms $\chi_{i0}^P$
$\chi_{i0}^f$ and $(\chi_{i0}^f)^2$, respectively). \par 

In the above formulae we have neglected the real parts of the reggeon exchanges given by
the signature factors. Since $\alpha_P(0)$ is not very different from unity, their effect is
expected to be rather small - except when the contribution of the multi-Pomeron exchanges
becomes very important. 
As for secondary ($f$) exchange, the signature factor enters only in terms with
multiple $f$-exchange - which are not important at the comparatively high 
energies of the present experiments. 
\par

Many parameters of the model such as $R_{0ki}$ are strongly constrained by the data on
hadronic interactions and were fixed (see Appendix 1). Let us also note that the $f$
exchange in $\sigma_{\gamma^*p}^{(tot)}$ is related to the valence quarks contribution,
which for small $x$ belongs to the $A$-component. Thus $C_S^f = 0$. \par

The list of all parameters and their numerical values is given in Appendix 1. 

\section{Diffractive production}
\hspace*{\parindent} Let us now introduce the expressions for diffraction dissociation of
a virtual photon. They can be obtained, using AGK-cutting rules, from the imaginary part of
the $\gamma^*p$-elastic scattering, given by eqs. (\ref{1e})-(\ref{8e}). 
Neglecting s-channel iterations of the triple-Pomeron graphs,
the total diffraction dissociation cross-section can be written as a sum of
three components

\beq
\label{9e}
\sigma_{\gamma^*p}^{(dif)} = \sum_{i=A,S} \sigma_i^{(0)} + \sigma_{PPP}
\eeq 

\noi where

\beq
\label{10e}
\sigma_i^{(0)} = 4 g_i^2(Q^2) \int \left ( \sigma_i^{(tot)}(b,s,Q^2) \right )^2 d^2b
\eeq  

\noi and

\beq
\label{11e}
\sigma_{PPP} = 2 \sum_i g_i^2(Q^2) \int \chi_{PPP}^i (b, s, Q^2) e^{-2C\chi_i(s,b,Q^2)} d^2b
\eeq

\noi where $\chi_{PPP}^i (b, s, Q^2) = a \chi_i^P(s, b, Q^2) \chi_3(s, b, Q^2)$, and
$\chi_i^P (s,b,Q^2)$ is given by the first term in the r. h. s. of 
eq. (\ref{5e}). For the total diffractive production
cross-section, which includes the diffraction dissociation of the proton, it is necessary
to multiply expressions (\ref{10e}), (\ref{11e}) by the factor $C = 1.5$, introduced in
Section~3. \par

Note that to first order in $a$, $\sigma_{PPP}$ corresponds to the sum of the 
two triple reggeon contributions in Fig. 5. Taken to all orders in $a$,
$\sigma_{PPP}$ corresponds to the resummation of all fan-type diagrams 
- obtained by adding to the diagrams in Fig. 5, multiple branchings 
with Pomeron and $f$-reggeon at each branching (see the discussion in 
Section 3).   
In the following, the sum of these two types of contributions will be referred
to as the triple Pomeron ($PPP$) - although the diagram in Fig. 5b corresponds
to an interference term. 
\par

With gaussian forms of the Born term in impact parameter, eq. (6), the 
expression (\ref{10e}) can be also written as follows (see second paper
of \cite{13r})

\beq
\label{12e}
\sigma_i^{(0)} = {g_i^2(Q^2) \sigma_{B_i} \over C} \left [ f\left ( {Z_i \over 2} \right )
- f(Z_i) \right ] \eeq 

\noi where

\beq
\label{13e}
f(Z) = \sum_{n=1}^{\infty} {(-Z)^{n-1} \over n \cdot n !}
\eeq

\beq
\label{14e}
\sigma_{B_i} = 2 \int \chi_i (s, b, Q^2) d^2b
\eeq

\beq
\label{15e}
Z_i = {8C \over \sigma_{B_i}} \int \chi_i^2(s,b,Q^2) d^2b \quad .
\eeq

\noi In the next sections we will compare our model with differential diffractive
cross-sections as functions of $M^2$ (square of the mass of the diffractively produced
system) or of $\beta = {Q^2 \over M^2 + Q^2}$ (which is convenient at large $Q^2$ and
plays the same role for the Pomeron as the variable $x$ for the proton\footnote{It
should be noted that the ``structure function of the Pomeron'' $F_P(\beta , Q^2)$ can be
used only for the cases when multi-Pomeron exchanges are small, because in general it is
impossible to separate out this amplitude.}) or of the variable $x_P$ ($x_p = {x \over
\beta}$). Present experiments at HERA measure the differential cross-sections integrated
over the variable $t$ (square of the momentum transfer between initial and final protons)
and usually the function $F_{2D}^{(3)}$ is introduced~:

\beq
\label{16e}
x_P \ F_{2D}^{(3)} = {Q^2 \over 4 \pi^2 \alpha_{e.m.}} \int x_P {d\sigma \over dx_Pdt} dt
\quad . \eeq 

\noi In our model it can be written as a sum of three terms 

\beq
\label{17e}
F_{2D}^{(3)} = \left ( \sum_i F_{2Di}^{(3)} (x, Q^2,\beta ) + F_{2DPPP}^{(3)}(x, Q^2,
\beta ) \right ) \eeq

\noi where

\beq
\label{18e}
F_{2Di}^{(3)} = F_{2Di}^{(3)B} \cdot \zeta_i \quad .
\eeq

\noi $F_{2Di}^{(3)B}(x,Q^2,\beta )$ are the lowest order (Born) approximations for these
functions (their explicit forms are given below) while $\zeta_i$ are ``suppression
factors'' due to higher order multi-Pomeron exchanges

\beq
\label{19e}
\zeta_i = \left [ f\left ( {Z_i \over 2} \right ) - f(Z_i) \right ] \cdot {8 \over Z_i}
\quad . \eeq

\noi The functions $F_{2Di}^{(3)}$ satisfy the condition

\beq
\label{20e}
\int_{x_{P_{min}}}^{x_{P_{max}}} F_{2Di}^{(3)} dx_P = \sigma_i^{(0)}(x, Q^2) \cdot {Q^2
\over 4 \pi^2 \alpha_{e.m.}}
 \eeq

\noi where $x_{P_{min}} = {x \over \beta_{max}}$~; $x_{P_{max}} = 0.1$~; $\beta_{max} = {Q^2
\over M^2_{min} + Q^2}$~: with $M^2_{min} = 4 m_{\pi}^2$. \par

The $\beta$-dependence of the $A$-component was chosen according to ref. \cite{15r}, where
the small-$\beta$-dependence was determined from the $PPf$-triple-Regge asymptotics with a
simple ansatz for the behaviour at $\beta \to 1$

\beq
\label{21e}
x_P\ F_{2DA}^{(3)B(P)}(x,Q^2,\beta ) = {Q^2 g_A^2(Q^2) \over 4 \pi^2 \alpha_{e.m.}} \int
d^2b \left (\chi_A^P   (x,Q^2,b) \right )^2 {\widetilde{\beta}^{2\Delta_P-\Delta_f} (1 -
\beta)^{n_P(Q^2)} \over \int_{\beta_{min}}^{\beta_{max}} {d\beta \over \beta}
\widetilde{\beta}^{2\Delta_p-\Delta_f}(1 - \beta )^{n_P(Q^2)}} \eeq

\noi where $\widetilde{\beta} = {Q^2 + S_0 \over Q^2 + M^2} = \beta \cdot {\widetilde{x}
\over x}$~; $\beta_{min} = {x \over x_P^{max}} = 10 x$. In the following all powers of $1- \beta$ are chosen according to the
arguments presented in refs. \cite{9r} and \cite{15r}: 

\beq
\label{22e}
n_P(Q^2) = - {1 \over 2} + {3 \over 2} \left ( {Q^2 \over c + Q^2} \right ) 
\eeq

\noi with $c = 3.5$ GeV$^2$. \par

The $S$-component should be maximal at $\beta$ close to 1 and was chosen in the form

\beq
\label{23e}
x_P\ F_{2DS}^{(3)B}(x, Q^2, \beta) = {Q^2 g_S^2 (Q^2) \over 4 \pi^2 \alpha_{e.m.}} \int d^2b\ 
\chi_S^2(x, Q^2, b) {\widetilde{\beta} \over \int_{\beta_{min}}^{\beta_{max}} {d\beta \over
\beta} \ \widetilde{\beta}} \quad . \eeq

\noi This component contains the contribution of comparatively small masses and should
decrease as $\beta \to 1$ faster than $PPf$ contribution, eq. (\ref{21e}). Our choice
satisfies this condition. Note that from perturbative QCD a behaviour $\beta^3$ has been
found \cite{20r}. However, the contribution of this component at large $Q^2$, where
perturbative QCD is applicable, is small $(\sim m^2_S/Q^2$). Our results are rather
insensitive to the power of $\beta$ in the range 1 to 3 \cite{22r}. 
\par

The triple-Pomeron ($PPP$ plus
$PfP$) contribution $F_{2DPPP}^{(3)}(x, Q^2, \beta )$, is given by 

\beq
\label{266}
x_P F_{2DPPP}^{(3)}(x, Q^2, \beta )
= x_P F_{2DPPP}^{(3)B}(x, Q^2, \beta )
{\sigma_{PPP} \over \sigma_{PPP}^B}
\quad , \eeq

\noindent where
$\sigma_{PPP}$ is given by eq. (11), its Born term, $\sigma_{PPP}^B$, by the
same equation with $C=0$, and

\beq
\label{24e1}
x_P F_{2DPPP}^{(3)B}(x, Q^2, \beta )= {Q^2 \over 4 \pi^2 \alpha_{e.m.
}} 2 a \int d^2b\
\sum_i g_i^2(Q^2) \chi_i^P(b,s,Q^2) \chi_3(s,b,Q^2,\beta ) \quad , \eeq

with

\beq
\label{24e}
\chi_3(s,b,Q^2,\beta ) = \sum_k \gamma_k \exp \left ( - {b^2 \over 4 \lambda_k\left (
{\widetilde{\beta} \over \tilde{x}}\right )} \right ) \left ( {\widetilde{\beta} \over
\widetilde{x}} \right )^{\Delta_k} {(1 - \beta )^{n_P(Q^2) + 4} \over \lambda_k\left (
{\widetilde{\beta} \over \tilde{x}}\right )} \quad .\eeq

\noi Here $\gamma_P = 1$, $\gamma_f$ determines the
relative strength of the $PfP$-contribution and $\lambda_k = R_{1k}^2 + \alpha '_K
\ell n \left ( {\widetilde{\beta} \over \tilde{x}}\right )$. The function
$\chi_3(s,b,Q^2)$, which enters in eqs. (\ref{5e}) and (\ref{11e}) is given by 

\beq
\label{25e}
\chi_3(s,b,Q^2) = \int_{\beta_{min}}^{\beta^{max}} {d \beta \over \beta} \ \chi_3(s,b,Q^2,
\beta ) \quad . \eeq    

\noi Since the triple Pomeron formula is not valid for low masses, we use here
$M_{min} = 1$~GeV.

Note that $x_P F_{2DPPP}^{(3)B}$ is obtained from $\sigma_{PPP}^B$ 
replacing $\chi_3(s,b,Q^2)$ by $\chi_3(s,b,Q^2,\beta)$.
In this way $\int dx_P F_{2DPPP}^{(3)}  dx_P
={Q^2 \over 4 \pi^2 \alpha_{e.m.
}} \sigma_{PPP}$.

\section{Comparison with experiment}
\hspace*{\parindent} The model outlined above was used for a joint fit of the data on the
structure function $F_2(x,Q^2)$ 
and diffractive structure function $F_{2D}^{(3)}(x,Q^2,\beta )$
in the region of small x ($x < 10^{-2}$) and 
$Q^2 \leq 10$ GeV$^2$\footnote{Actually, for $F_2$ only data with 
$Q^2 \leq 3.5$ GeV$^2$ were included in the fit.}. 
The full list of parameters and their values either fixed or obtained in the fit are given
in the Appendix 1.
\par

In Fig.~6, the dependence of $\sigma_{\gamma^*p}$ as a function of $Q^2$ for different
energies is shown and Fig.~7 gives the $x$-dependence of $F_2(x, Q^2)$ for different
$Q^2$. Our results are compared with experimental data. The 
description of the data
is excellent. Note that the $x$-dependence and its variation with $Q^2$ is fixed in the
model and is strongly correlated with the ratio of
$\sigma_{\gamma^*p}^{(dif)}/\sigma_{\gamma^*p}^{(tot)}$ and its dependence on $Q^2$. The
model confirms the $Q^2$-dependence of the effective intercept $\Delta_{eff} = {d \ell n
F_2(x,Q^2) \over d \ell n \left ( {1 \over x}\right )}$, assumed in ref. \cite{9r}. The
function $\Delta_{eff}$ depends not only on $Q^2$, but also on $x$ and it is shown in Fig.~8
for two different intervals of $x$. The function $\Delta_{eff}$ decreases as $Q^2$ or $x$
decreases due to the increase of shadowing effects. It is interesting to note that, for
$Q^2 = 0$, $\Delta_{eff}$ in the region of HERA energies is close to 0.13, i.e.
substantially higher than in hadronic interactions. For $s \to \infty$ the cross-section
$\sigma_{\gamma^*p}^{(tot)}$ has a Froissard-type behavior in $\ell n^2 {s \over s_0}$.
\par

In Figs.~9 and 10 the results of the model are compared with experimental data 
on
diffractive structure function $F_2^{D(3)}(x, \beta , Q^2)$. The $\beta$
dependence for fixed values of $Q^2$ and $x_P$ is shown in Fig.~9 and the
$x_P$-dependence for different values of $Q^2$ and $\beta$ is presented in Fig.~10. The
model reproduces the experimental data rather well. For comparison we limited ourselves to
the data at low values of $x_P \leq 10^{-2}$, where the effect of the non-diffractive RRP
contribution (which is not included in the model) is small. Diffractive production at
$Q^2 = 0$ is reasonably well reproduced as well (Fig.~11). The dependence of diffractive
cross-sections on $\Delta_{eff}^{dif} \left ( F_{2D}^{(3)} \sim \left ( {1 \over x_P} \right
)^{2\Delta_{eff}^D}\right )$ is shown in Fig.~12 as a function of $Q^2$ 
(for fixed $M$).
The experimental points are also shown. The model predicts a weak dependence of
$\Delta_{eff}^{dif}$ on $Q^2$ for $Q^2 > 5 \div 10$~GeV$^2$ and QCD-evolution does not
change this result \cite{15r}. 

\section{Discussion and conclusions}
\hspace*{\parindent} The model we propose here for the description of total and
diffractive cross-sections of interaction of a virtual photon with a proton is a natural
generalization of models used for the description of high-energy hadronic interactions.
The main parameter of the model - intercept of the Pomeron, $\Delta_P \equiv \alpha_P(0) -
1$, was fixed from a phenomenological study of these interactions ($\Delta_P = 0.2$) and
was found to give a good description of $\gamma^*p$-interactions in a broad range of
$Q^2$ ($0\le Q^2< 10\  GeV^2$). 
Note that a single Pomeron is present in the model. \par

It should be noticed that at higher values of $Q^2$, QCD evolution is important and
our results should be 
used as initial
condition in DGLAP evolution (see refs \cite{11r,15r}). 
In particular, due to QCD evolution, the effective Pomeron
intercept in $F_2$ will reach values significantly larger than 1.2 at large $Q^2$. For diffractive
process the situation is different. In this case, 
the effective Pomeron intercept at large $Q^2$ is not
significantly modified as a result of QCD evolution. Moreover, the latter has rather small
effect for intermediate values of $\beta$ \cite{15r}. For these reasons our model, without
QCD evolution, can be used here at comparatively larger values of $Q^2$. \par

Another important parameter of the theory is the triple Pomeron vertex, 
which appears in the parameter $a$. It is quite remarkable that 
the same value of this parameter allows to describe diffractive data both at 
$Q^2=0$ and at a few GeV$^2$. Thus, in our approach, the value of this 
parameter can be determined from soft diffraction data \cite{25r}.
\par

 The variation of the photon virtuality gives a unique
possibility to study unitarization effects for different scales. Let us discuss the
qualitative features of our model and its relation to the models introduced by other
authors \cite{2r}. It has been mentioned above that the different role of small and
large size components of a virtual photon has been emphasized by many authors \cite{2r} and
perturbative calculations of unitarization effects for $\sigma_{\gamma^*p}$ have been
carried out in ref. \cite{2r}. In our opinion the perturbative calculations can be
valid only for the small size component ($S$) at large $Q^2$. An interesting analysis of
unitarization effects for both total $\gamma^*p$-cross-section and diffractive production
was performed 
by Golec-Biernat and W\"usthoff
\cite{2r}. These authors, however, postulated an unconventional
form for the elastic $\gamma^*p$-amplitude and it is not clear which diagrams lead to these
unitarization effects\footnote{Correlations in hadronic interactions have been
studied for many years. It is known that they can be described 
\protect{\cite{27r}}
in an eikonal model - where the Born term, at fixed impact parameter, is used as
eikonal. In the model of Golec-Biernat and W\"usthoff \protect{\cite{2r}}, on the contrary, the Born term
is integrated over impact parameter and then eikonalized. The unitarity effects
obtained in this way are very large. It is unlikely that they can describe the
observed features of correlations.}. We agree with their conclusion 
that for the structure function $F_2(x,Q^2)$ the unitarization effects at
HERA energies are important mostly in the region of rather small $Q^2 \ \lsim \
2 \div 4$~GeV$^2$. 
Indeed, for $Q^2 > 2 \div 4$~GeV$^2$, the experimental value of
$\Delta_{eff}$ is close to the one of the input (bare) Pomeron 
$\Delta_P = 0.2$.
Within our approach this is only possible if the $S$-component (where unitarity
corrections are higher twists) dominates. However, some unitarity corrections are
present at large $Q^2$ due to the $A$-component and the triple Pomeron interaction.
Moreover, they will become more important at higher energies (see the discussion in
the next paragraph). However, we differ in many details from their predictions 
on the properties of diffractive production at large $Q^2$ and on the pattern of
``saturation'' of parton densities at large $Q^2$ and extremely small $x$. \par

Let us now discuss the saturation patterns in our model, i.e. the properties of
$\sigma_{\gamma^*p}^{(tot)}(s,Q^2)$ in the limit $s \to \infty$. It follows from formulae
(\ref{1e})-(\ref{8e}) that for the $A$-component, which has a large cross-section,
$\sigma_{\gamma^*p}^{(tot)}(s,b,Q^2)$ will tend to a saturation limit $1/2C$ rather fast
as energy increases. On the other hand, for the small size (at large $Q^2$) component
($S$) the situation is different. If we neglect the triple-Pomeron contribution (i.e. with
$a = 0$ in eq. (\ref{5e})), $\chi_S(s,b,Q^2) \sim {Cm_S^2 \over Q^2} \exp \left (
\Delta_P\xi - {b^2 \over 4 \lambda_{0P}^S} \right )$ will increase with energy until the
increase of $\exp (\Delta_P\xi )$ does not overcompensate the smallness of ${m_S^2 \over
Q^2}$. For such extremely large energies, cross-sections in the small impact parameter
region will saturate to a $Q^2$-independent value and $F_2(x,Q^2) \sim Q^2$. This is a
usual picture of saturation in perturbative QCD \cite{1r}-\cite{3r}. However, the inclusion
of the $PPP$-fan diagrams (large distance, nonperturbative effects) according to eqs.
(\ref{5e}), (\ref{24e}), (\ref{25e}) changes the picture drastically. In this case the
increase of $\chi_3(s,b,Q^2) \sim \exp (\Delta \xi )$ will compensate the
increase of $\chi_S(s,b,Q^2)$ at very large $\xi$ and lead, at large $Q^2$, to a behaviour
$\sigma_{\gamma^*p}^{(tot)} \sim {1 \over Q^2} f(\ell n Q^2)$ even for $x \to 0$. (Note
that for large enough $Q^2$ this will happen in a region where $\chi_S$ is still small).
\par

The above effects are relatively small at present energies due to the smallness of the
triple Pomeron vertex $r_{PPP}$ which determines the constant $a$. However, in the
``saturation limit'' the properties of $\sigma_{\gamma^*p}^{(tot)}(s, Q^2)$ in our model
are quite different from those in other models \cite{2r,3r}. Our predictions for
$\sigma_{\gamma^*p}^{(tot)}(s,Q^2)$ for energies higher than those accessible at HERA are
shown in Figs.~13. \par

In a forthcoming publication \cite{22r} we introduce a different but closely related model
with also two components for the interaction of the $q\bar{q}$ pair~: a large-size
component, parametrized in the same way as the $A$-component in the present work, and a
small-size component computed using the $q\bar{q}$ perturbative QCD wave function - with its
longitudinal and transverse components. The results are very similar to the ones obtained
here. Again, a single Pomeron with intercept 1.2 allows to describe the data on both total
$\gamma^*p$ cross-section and diffractive production.  \par

\newpage
\section*{Appendix 1}
\hspace*{\parindent} In this Appendix we give the list of all the parameters in the
formulae for $\sigma_{\gamma^*p}^{(tot)}(s,Q^2)$ (or $F_2(x, Q^2)$) and diffractive
production, given in sections 3 and 4 res\-pec\-ti\-ve\-ly - together with their numerical
values. The total cross-section of $\gamma^*p$ interactions is related to the structure
function $F_2$ as follows

$$\sigma_{\gamma^*p}^{(tot)}(s, Q^2) = {4 \pi^2\alpha_{em} \over Q^2} \ F_2(x,Q^2) \quad ,
\eqno({\rm A.1})$$

\noi where $s = W^2$ and $x = {Q^2 \over 1 + Q^2}$. This cross-section is related by eq.
(\ref{1e}) to the corresponding quantity, $\sigma_{\gamma^*p}^{(tot)}(b,s,Q^2)$, in the
impact parameter space. The latter is written as the sum of two components (eq.
(\ref{2e})) with probabilities given eq. (\ref{3e}) where $g_A(0)$, $g_S(0)$ and $m_A^2$
are free parameters. Their values obtained from the fit are

$$g_A(0)= 0.15 \alpha_{e.m.} \qquad , \qquad  g_S(0)= 0.367 \alpha_{e.m.}
 \qquad , \qquad  
m_A^2= 0.227 \ {\rm GeV}^{2} \quad . \eqno({\rm A.2})$$ 

For $a = 0$ (i.e. when triple reggeon vertices vanish~; see eq. (\ref{5e})) the
cross-sections $\sigma_i^{(tot)}(b,s, Q^2)$ are written in a quasi-eikonal form (eq.
(\ref{4e})) with Born terms given by the exchange of the Pomeron $P$ and secondary
reggeon ($f$) trajectories). Their expressions contain the intercepts $\alpha_k(0) =
1 + \Delta_k$ $(k = P, f)$ for which we take

$$\Delta_P = 0.2 \qquad , \qquad \Delta_f = - 0.3 \quad . \eqno({\rm A.3})$$

\noi They also contain five other parameters treated as free ones~: $C_A^P$, $C_S^P$,
$C_A^f$, ($C_S^f = 0$), $s_0$ and $m_S^2$. Their values obtained in the fit 
are:

$$ m_S^2= 1.343 \ {\rm GeV}^{2} \qquad , s_0= 0.463 \ {\rm GeV}^{2} \qquad , $$
$$C_A^P= 0.830 \ {\rm GeV}^{-2} \qquad , C_S^P= 0.807 \ {\rm GeV}^{-2} \qquad ,
C_A^f= 14.28 \ {\rm GeV}^{-2} \quad . \eqno({\rm A.4})$$
\par

The value of the Pomeron intercept $\alpha_P(0) = 1.2$ has been found in soft hadronic
interactions \cite{8r}. It is also consistent with the intercept of the BFKL Pomeron at NLO.
When treated as a free parameter in the fit its value turns out to be very close to the one
in eq. (A.3). The value of the intercept of the $f$ has been allowed to change within
some restricted limits. The value $\alpha_f(0) = 0.7$ leads to better results than the
more conventional value $\alpha_f(0) = 0.5$. \par

Our expressions contain also the triple Pomeron term $PPP$ whose strength is 
controlled by
the parameter $a$ in eq. (\ref{5e}), and the interference term $PfP$ whose strength
(relative to the $PPP$ term) is given by the parameter $\gamma_f$. We use

$$a = 0.052 \ 
{\rm GeV^{-2}}
\qquad \gamma_f = 8 \quad . \eqno({\rm A.5})$$    

\noi Note that the value of $a$ 
obtained in the fit agrees with the value obtained
from soft diffraction (see the discussion in Section 6). \par

Finally we turn to the quantities

$$\lambda_{0k}^i = R_{0ki}^2(Q^2) + \alpha '_k \xi \eqno({\rm A.6})$$

\noi and 

$$\lambda_k = R_{1k}^2 + \alpha '_k \ell n (\beta /x ) \eqno({\rm A.7})$$

\noi in eqs. (\ref{5e}) and (\ref{24e}). They are directly related to the 
$t$-dependence
of the elastic $\gamma^*p$ amplitudes and of the triple reggeon vertices, respectively,
and have the standard regge behavior as functions of $\xi$ - or $\beta /x$. We use the
conventional parameters of the reggeon slopes

$$\alpha '_P = 0.25 \ {\rm GeV}^{-2} \qquad , \qquad \alpha '_f = 0.9 \ {\rm GeV}^{-2} \quad
. \eqno({\rm A.8})$$

Our results for the diffractive cross-sections integrated over $t$ are not sensitive to
the values of the parameters $\lambda$ and we have fixed them to the values obtained from
soft interaction data and from vector dominance. We take \cite{28r}

$$\begin{array}{l} R_{0kA}^2(Q^2) = R_{vkA}^2 + R_{pk}^2 \\ \\ R_{0kS}^2(Q^2) =
\displaystyle{{R_{vkS}^2 \over 1 + {Q^2 \over m_{\rho}^2}}} + R_{pk}^2 \end{array}
\eqno({\rm A.9})$$ 

\noi with

$$R_{pP}^2 = R_{Pf}^2 = 2 \ {\rm GeV}^{-2} \qquad , 
\qquad R_{vki}^2 = 1 \ {\rm GeV}^{-2} 
\eqno({\rm A.10})$$

\noi and

$$R_{1P}^2 = R_{1f}^2 = 2.2 \ {\rm GeV}^{-2} \quad . \eqno({\rm A.11})$$

\noi Note that the $Q^2$-dependence of the $\gamma^*\gamma^*P$ and $\gamma^*\gamma^*f$ 
vertices for the $S$ component have been chosen in such a way that the corresponding
radii tend to zero for $Q^2 \to \infty$ \cite{28r}. \\

\noi {\bf Acknowledgments} \\

It is a pleasure to thank K. Boreskov, O. Kancheli, G. Korchemski, U. Maor and C. Merino
for discussions. This work is partially supported by NATO grant OUTR.LG 971390.
E. G. F. and C. A. S. thank Ministerio de Educaci\'on y
Cultura of Spain for financial support.

%
%

\newpage

\section*{Figure captions}

\vskip 0.5cm

\noindent{\bf Figure 1.} 
Single Pomeron exchange (Fig.~1a) and multi-Pomeron
exchanges (Figs.~1 b, 1c).

\vskip 0.5cm

\noindent{\bf Figure 2.} 
Intermediate
states that can be diffractively produced by a single Pomeron exchange.

\vskip 0.5cm

\noindent{\bf Figure 3.} 
Diagram for diffractive vector mesons production (Fig.~3a) and
triple-Regge diagram (Fig.~3b) with $P$ and
$f$-exchanges in the $t$-channel ($PPP$ and $PPf$).

\vskip 0.5cm

\noindent{\bf Figure 4.} 
Fan-type diagrams with Pomeron branchings.

\vskip 0.5cm

\noindent{\bf Figure 5.} 
Triple Pomeron diagrams: a) triple Pomeron $PPP$ - already represented in Fig. 
3b; b) the interference term $PfP$.

\vskip 0.5cm

\noindent{\bf Figure 6.} The $\gamma^{*} p$ cross section as a function of
$Q^2$ for different energies compared to the following experimental data:
H1 1995 \cite{a} (black points), ZEUS 1995 \cite{b} (black squares),
E665 \cite{c}
(black triangles) and ZEUSBPT97 \cite{d} (white circles). The data at
$Q^2=6.5$ GeV$^2$ (H1 1994 -black stars- and ZEUS 1994 -white stars-)
are from \cite{ab}.
\vskip 0.5cm

\noindent{\bf Figure 7.} $F_2$ as a function of $x$ for different values of $Q^2$.
The experimental data are the same of Fig. 6.
\vskip 0.5cm

\noindent{\bf Figure 8.} The function $\Delta_{eff}$ as a function of $Q^2$ for 2 intervals
of $x$. The solid line corresponds to the interval 
$\widetilde{x}=\frac{Q^2+s_0}{Q^2+s}=1.85*10^{-4} \div
1.16*10^{-5}$
and the dashed line to the interval 
$\widetilde{x}=7.41 *10^{-14} \div 
4.64*10^{-15}$.
\vskip 0.5cm

\noindent{\bf Figure 9.} The diffractive structure function $x_P F_2^{D(3)}$
as a function of $\beta$ for fixed values of $Q^2$ and $x_P$.
The experimental data (H1 1994) are from \cite{e}.
\vskip 0.5cm

\noindent{\bf Figure 10.} The diffractive structure function $x_P F_2^{D(3)}$
as a function of $x_P$ for fixed values of $Q^2$ and $\beta$.
The experimental data (H1 1994 and H1 1995)
are from \cite{e} and \cite{f}.
\vskip 0.5cm

\noindent{\bf Figure 11.} Diffractive production at $Q^2=0$ GeV for two different energies.
The experimental data are from \cite{g}.
\vskip 0.5cm

\noindent{\bf Figure 12.} The effective Pomeron slope $\Delta^{D}_{eff}$ as a function of
$Q^2$  for $M_X=5$ GeV. The experimental values are from ZEUS 1994
\cite{h}.
\vskip 0.5cm

\noindent{\bf Figure 13.} Our predictions for
the $\gamma^{*} p$ cross section at
 energies higher than those accessible at HERA.
The experimental data are the same as in Figs. 6 and 7.

\newpage

%
%

\centerline{\bf Figure 1}
\vspace{1cm}

\hspace{-1.2cm}\epsfig{file=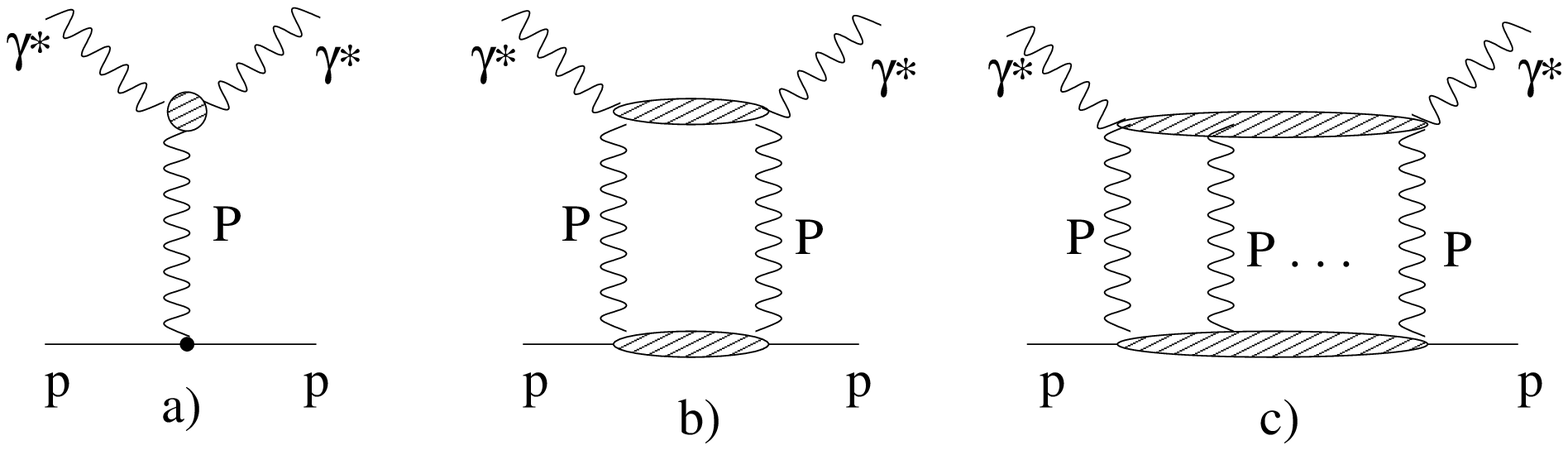,width=16.cm}

\newpage

\centerline{\bf Figure 2}
\vspace{1cm}

\hspace{-1.2cm}\epsfig{file=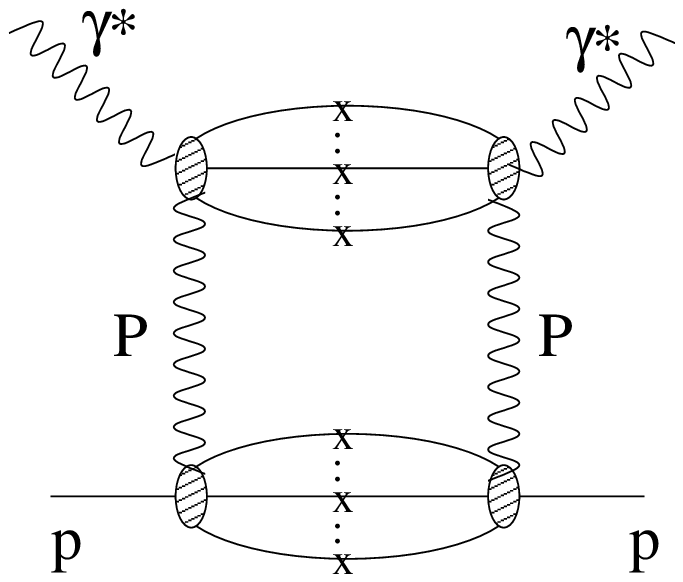,width=7.cm}

\newpage

\centerline{\bf Figure 3}
\vspace{1cm}

\hspace{-1.2cm}\epsfig{file=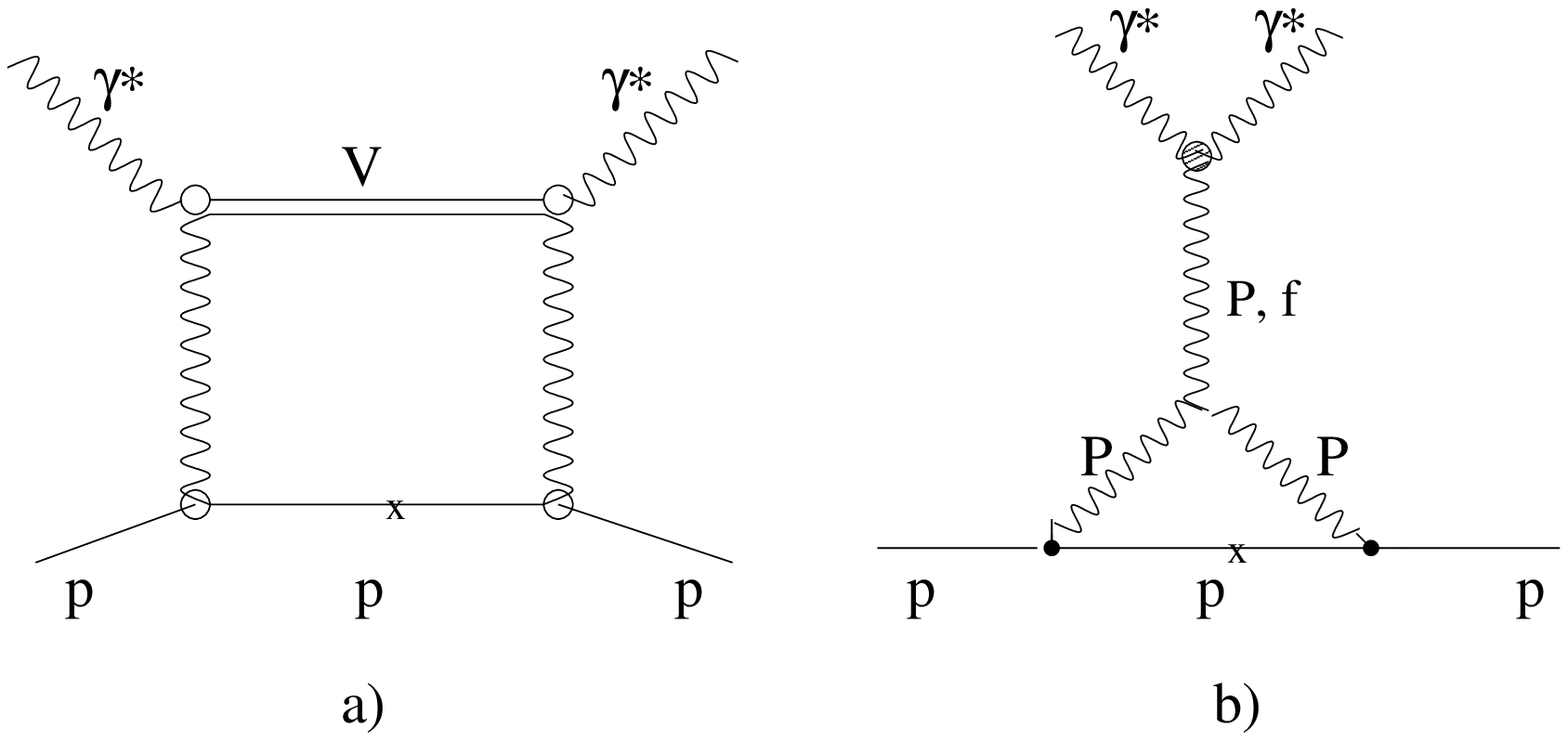,width=16.cm}

\newpage

\centerline{\bf Figure 4}
\vspace{1cm}

\hspace{-1.2cm}\epsfig{file=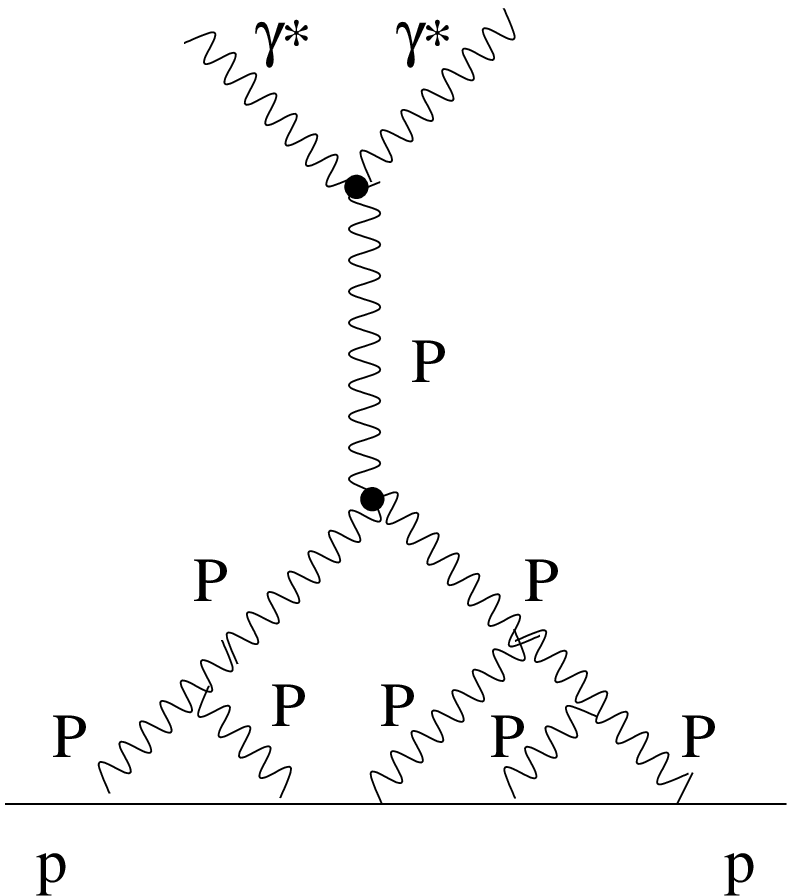,width=10.cm}

\newpage

\centerline{\bf Figure 5}
\vspace{1cm}

\hspace{-1.2cm}\epsfig{file=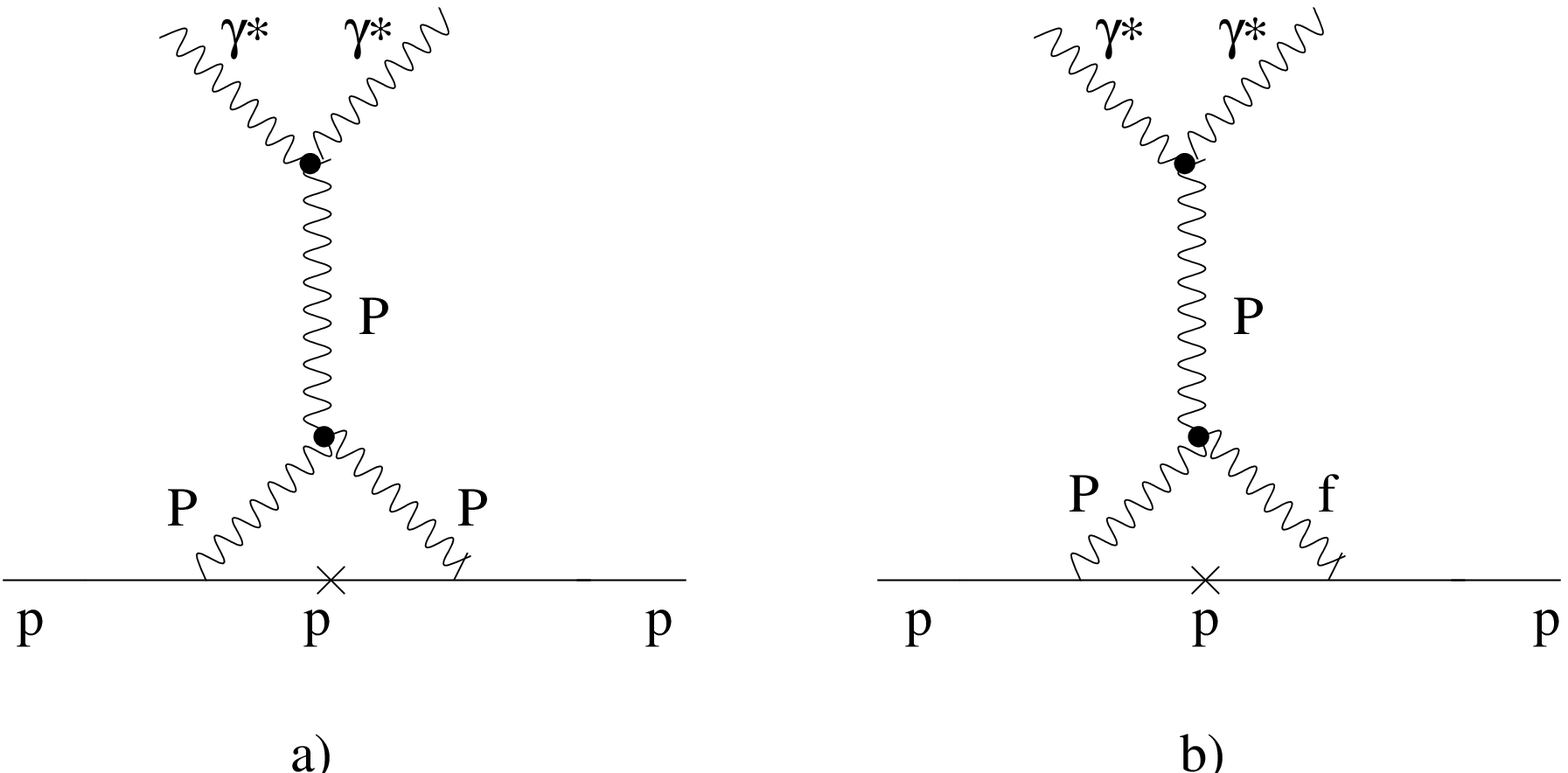,width=20.cm}

\newpage

\centerline{\bf Figure 6}

\vspace{1cm}

\hspace{-1.2cm}\epsfig{file=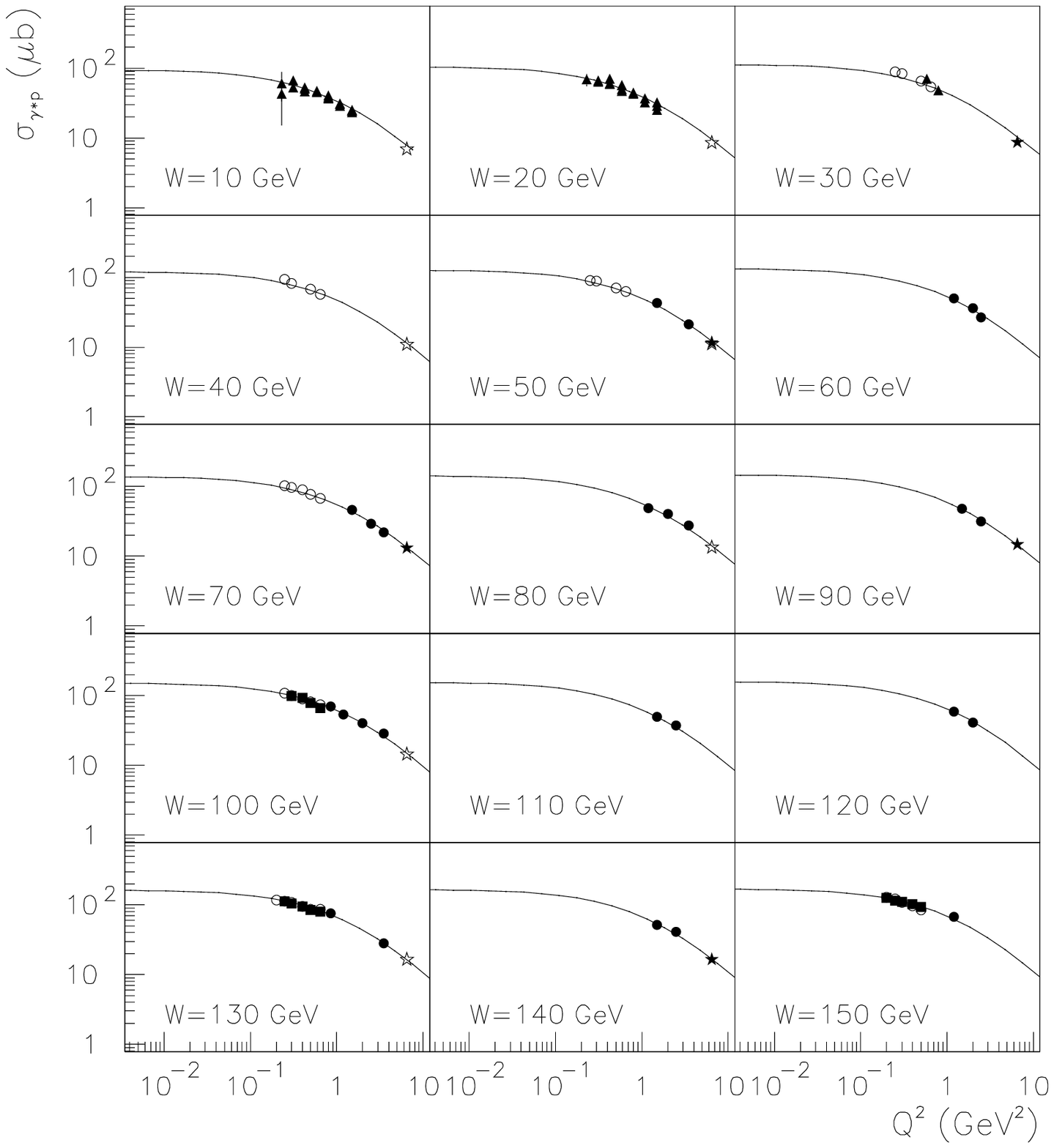,width=17.cm}

\newpage

\vspace{1cm}

\hspace{-1.2cm}\epsfig{file=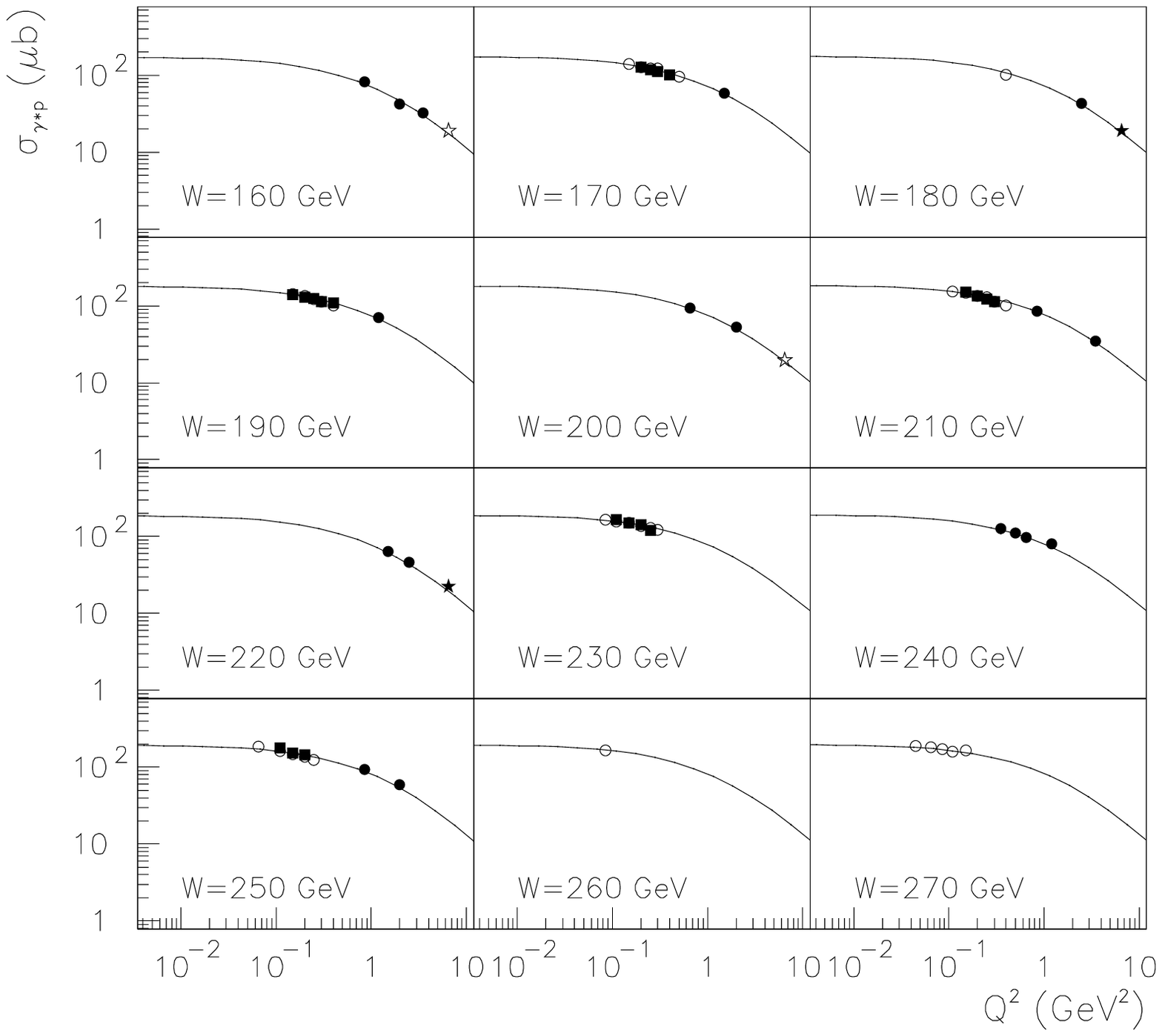,width=17.cm}

\newpage

\centerline{\bf Figure 7}
\vspace{1cm}

\hspace{-1.2cm}\epsfig{file=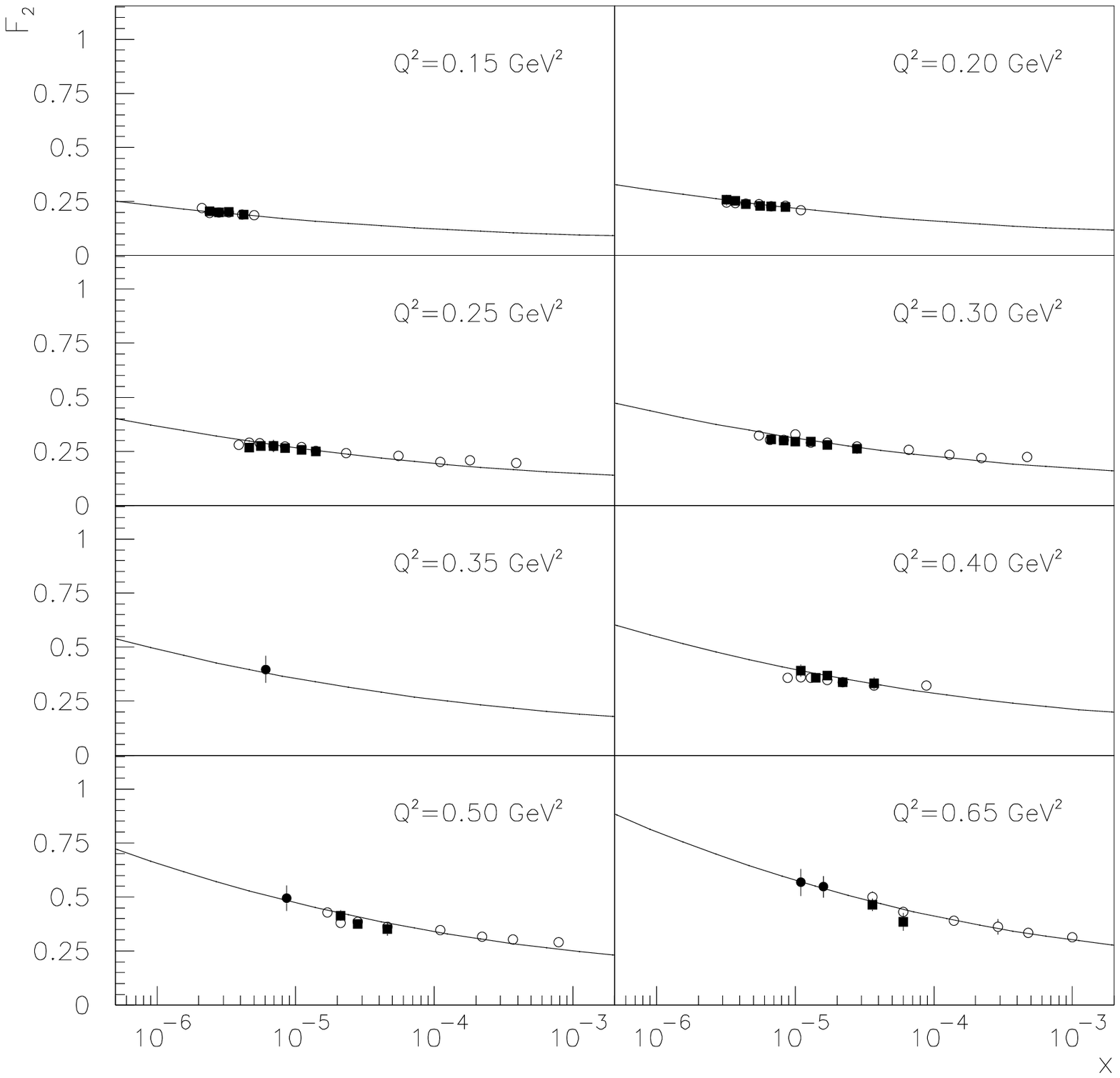,width=17.cm}

\newpage

\vspace{1cm}

\hspace{-1.2cm}\epsfig{file=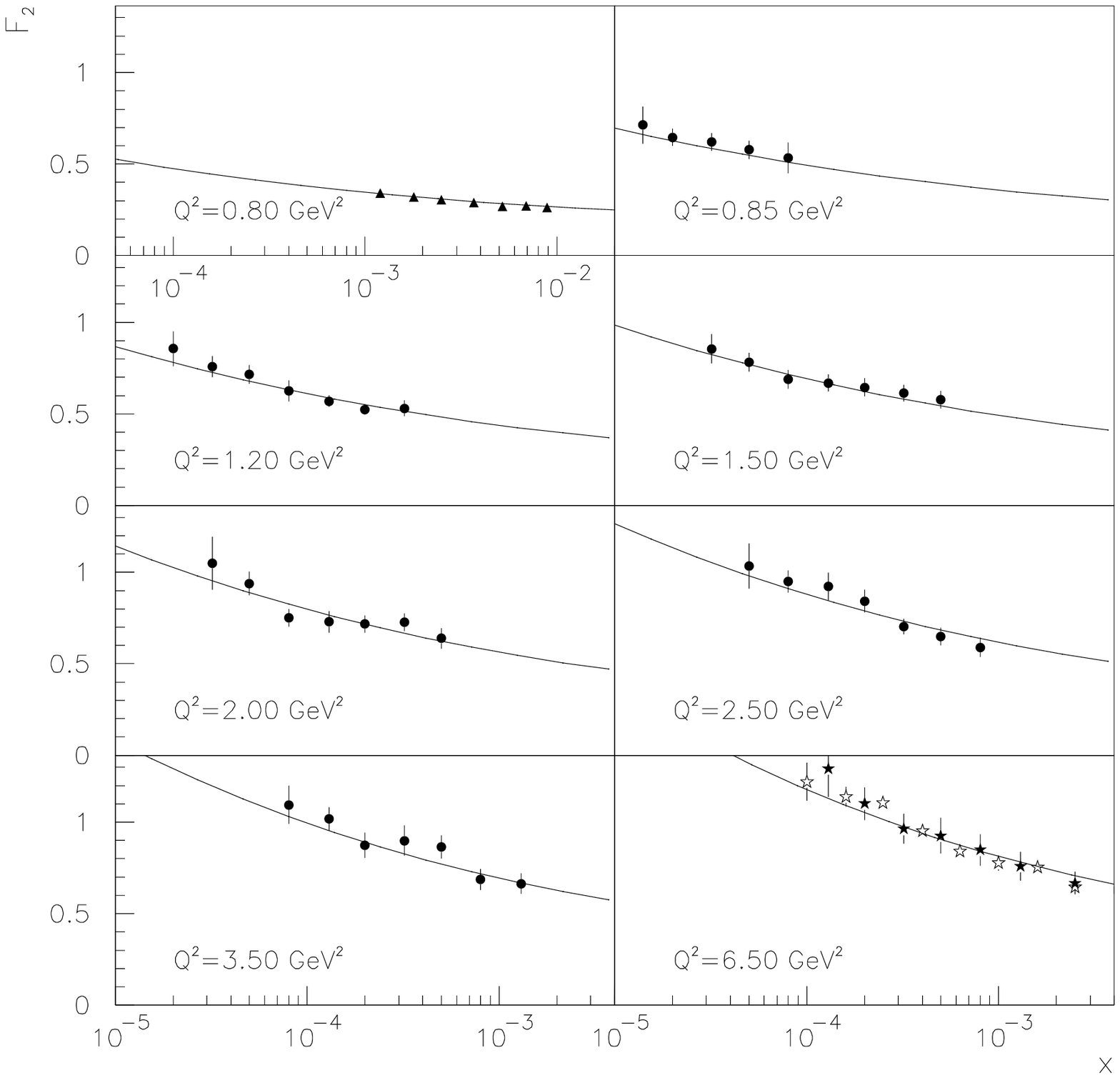,width=17.cm}

\newpage

\centerline{\bf Figure 8}
\vspace{1cm}

\hspace{-1.2cm}\epsfig{file=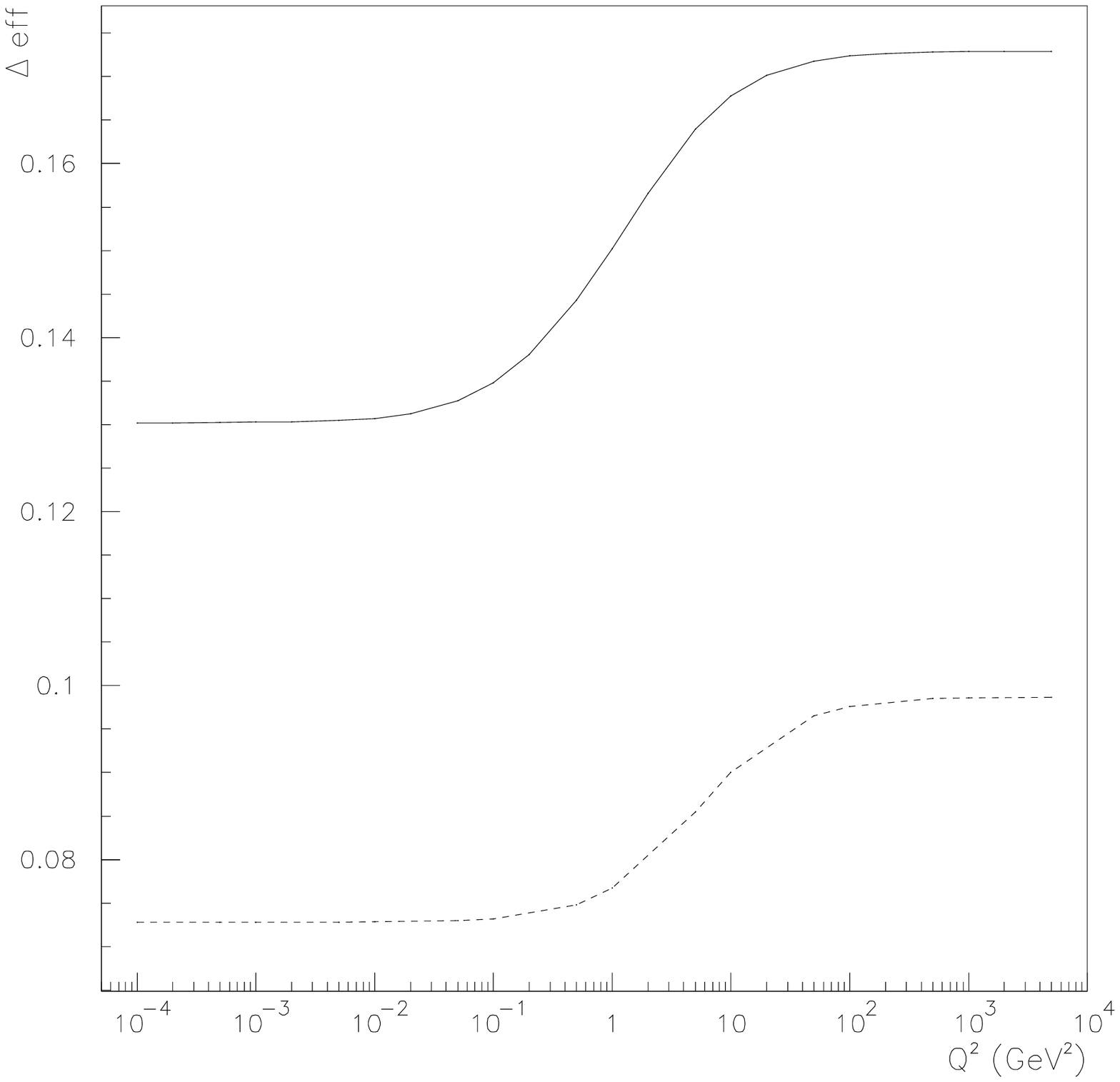,width=17.cm}

\newpage

\centerline{\bf Figure 9}
\vspace{1cm}

\hspace{-1.2cm}\epsfig{file=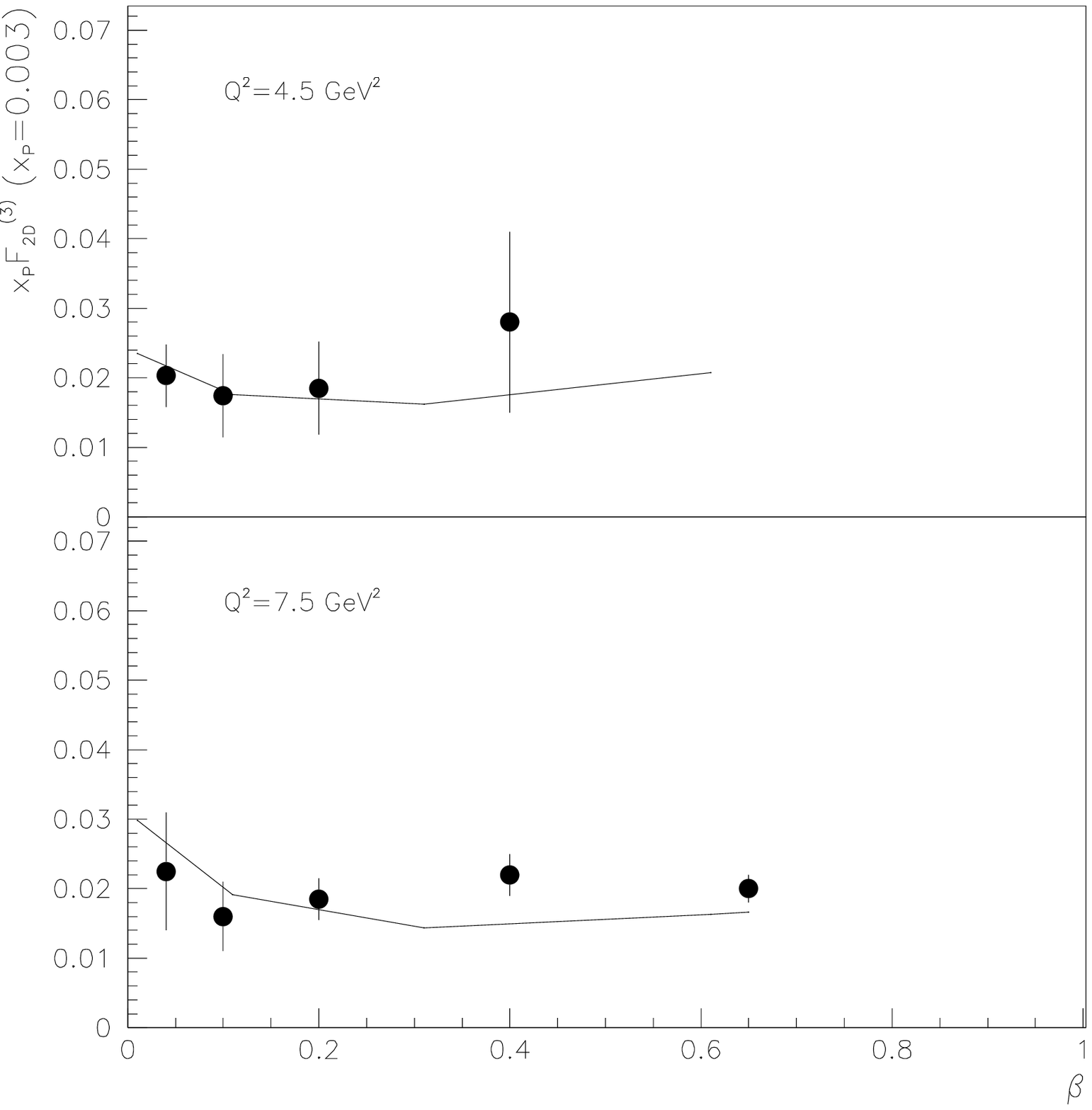,width=17.cm}

\newpage

\centerline{\bf Figure 10}
\vspace{1cm}

\hspace{-1.2cm}\epsfig{file=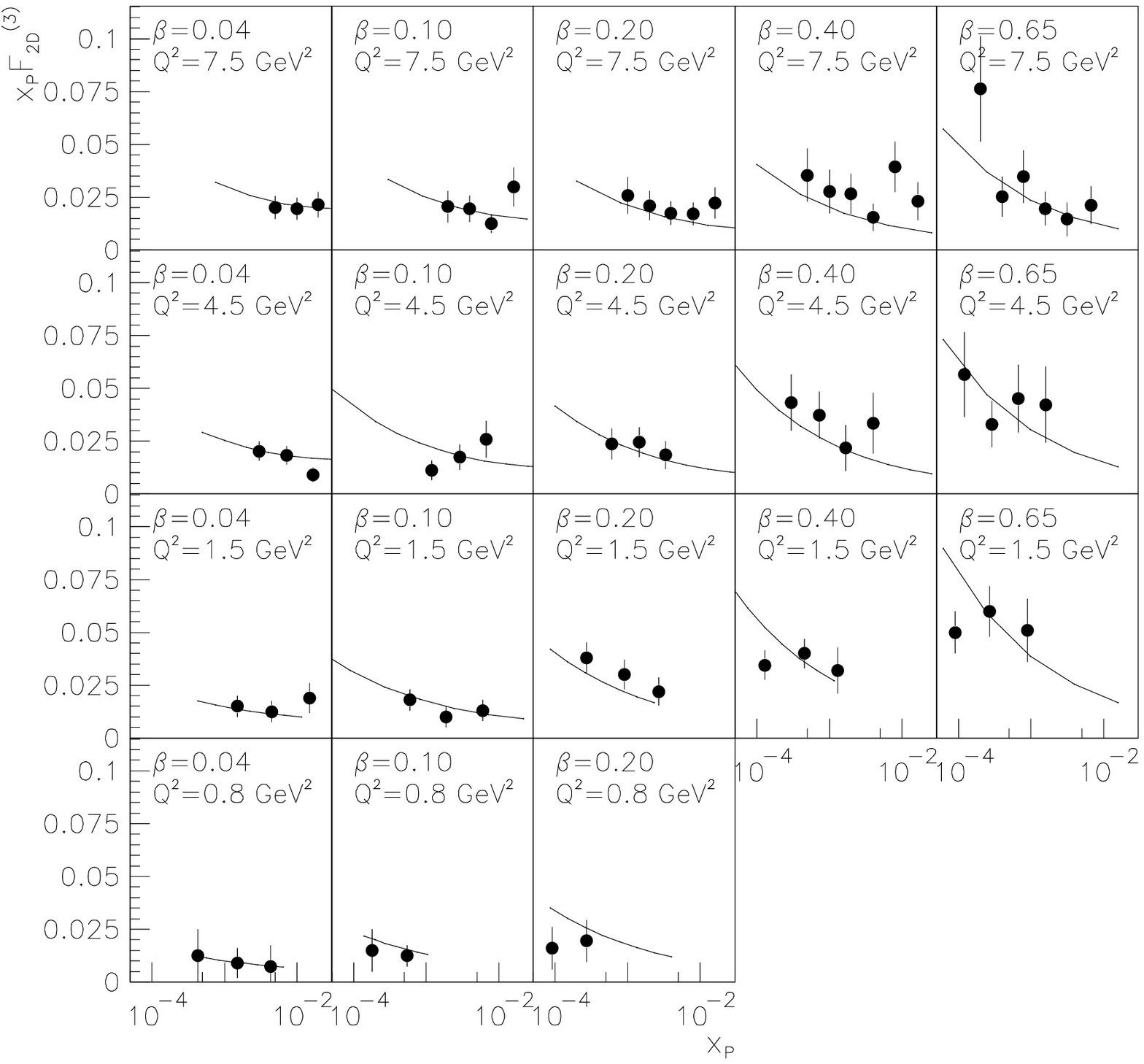,width=17.cm}

\newpage

\centerline{\bf Figure 11}
\vspace{1cm}

\hspace{-1.2cm}\epsfig{file=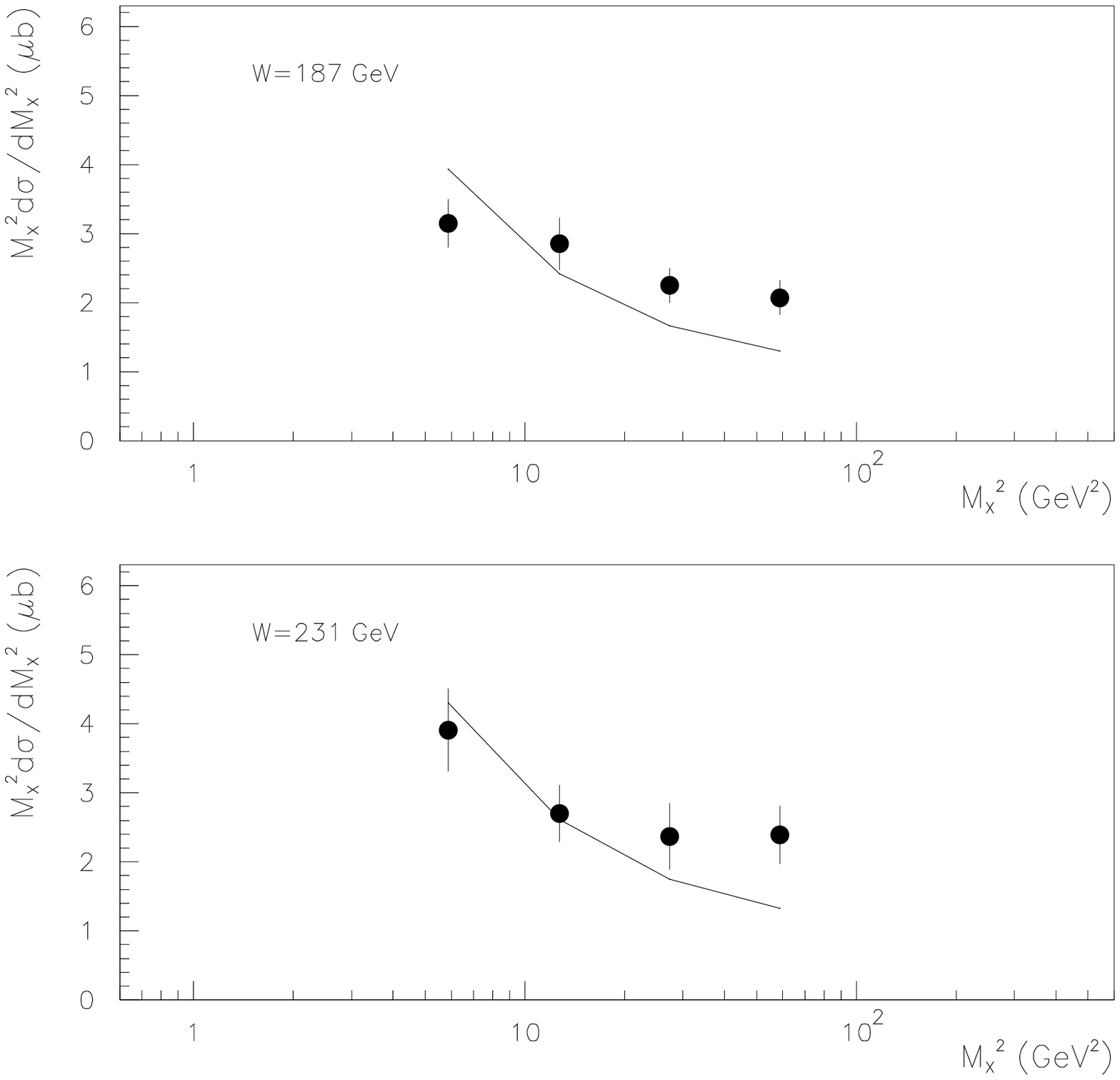,width=17.cm}

\newpage

\centerline{\bf Figure 12}
\vspace{1cm}

\hspace{-1.2cm}\epsfig{file=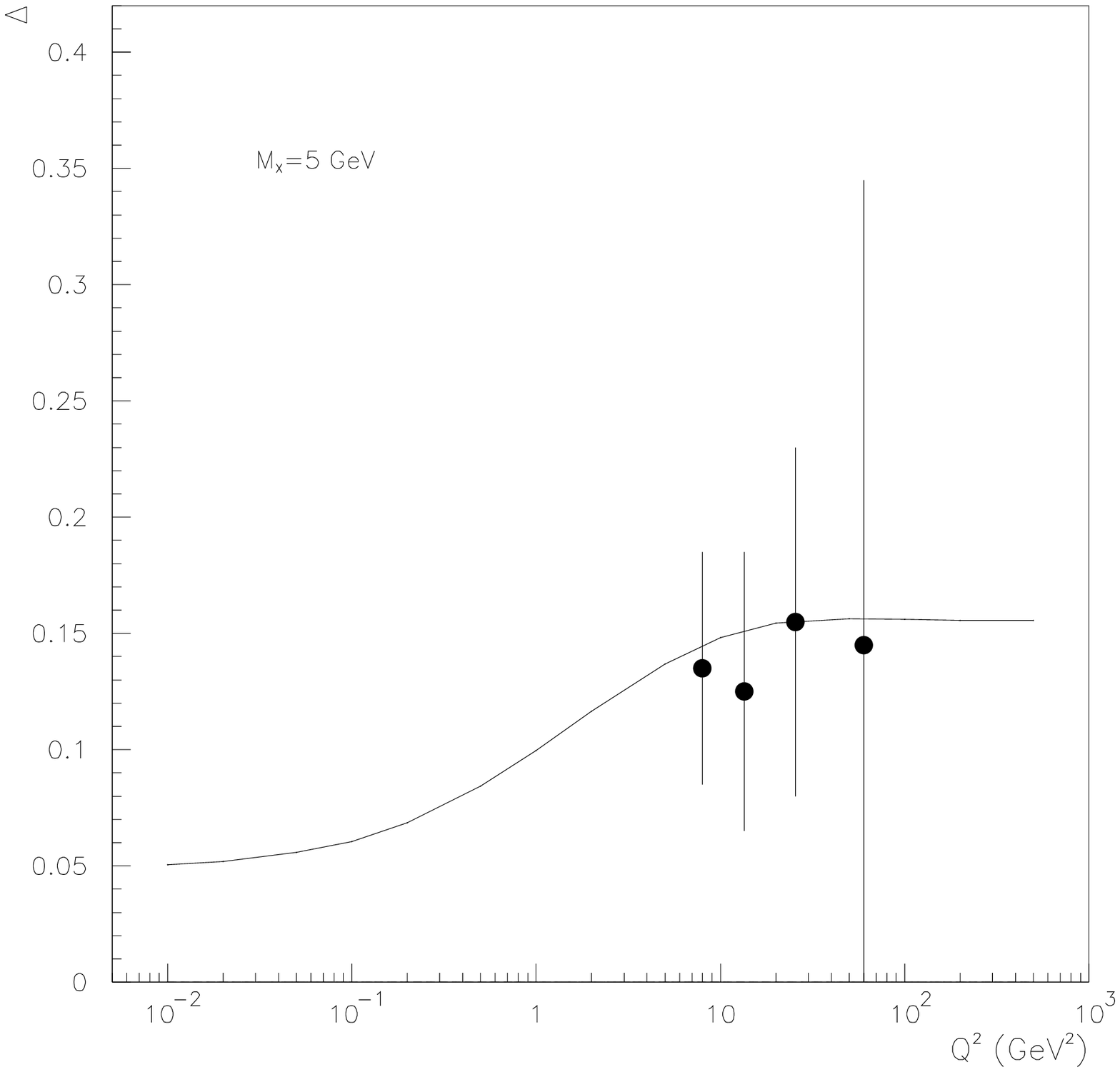,width=17.cm}

\newpage

\centerline{\bf Figure 13}
\vspace{1cm}

\hspace{-1.2cm}\epsfig{file=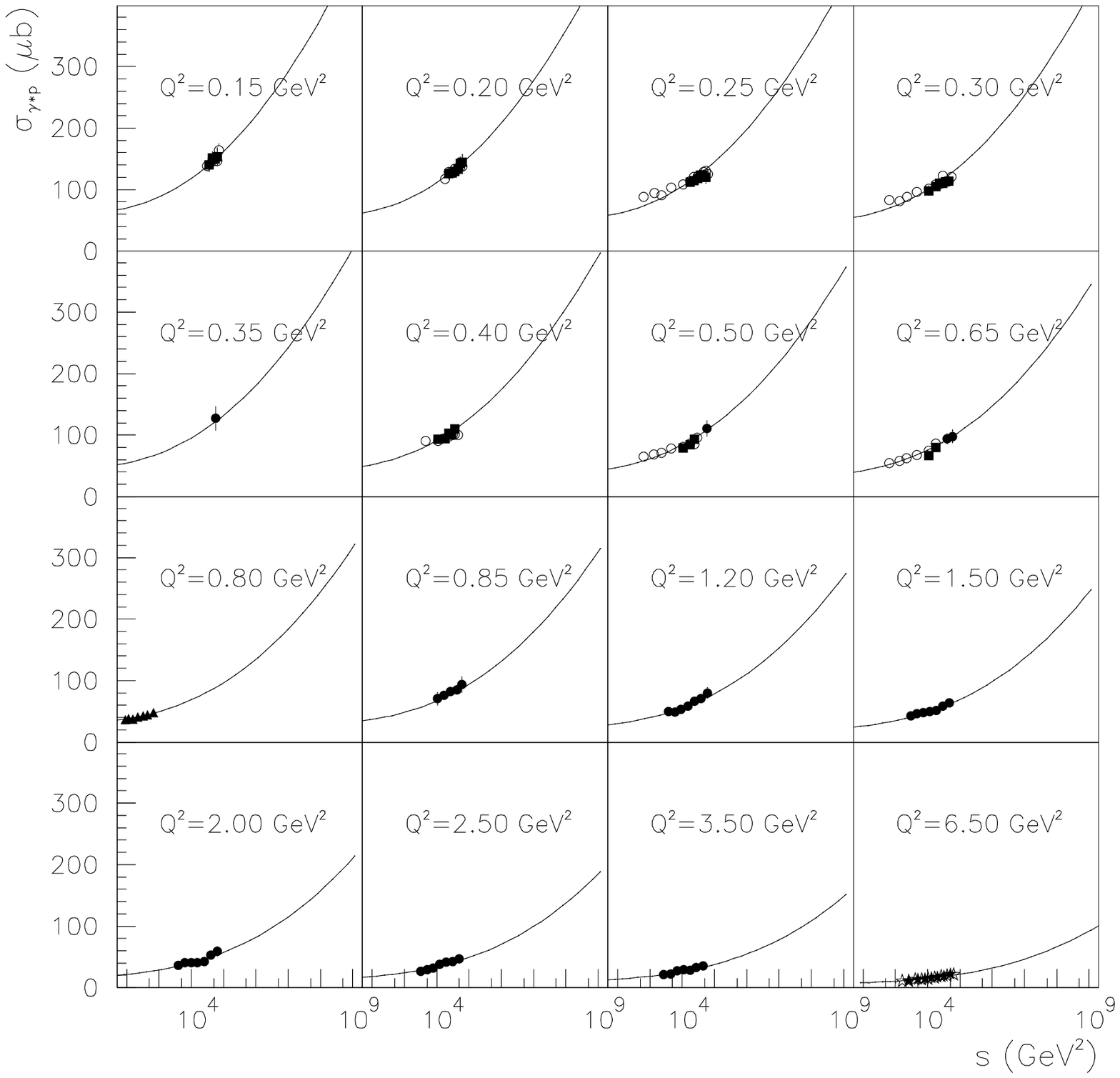,width=17.cm}

\end{document}